\documentclass[sigconf]{acmart}

\copyrightyear{2026}
\acmYear{2026}
\acmConference[TEI '26]{Twentieth International Conference on Tangible, Embedded, and Embodied Interaction}{March 08--11, 2026}{Chicago, IL, USA}
\acmBooktitle{Twentieth International Conference on Tangible, Embedded, and Embodied Interaction (TEI '26), March 08--11, 2026, Chicago, IL, USA}
\acmDOI{10.1145/3731459.3773336}
\acmISBN{979-8-4007-1868-7/26/03}

\usepackage[utf8]{inputenc}
\usepackage{subcaption}
\usepackage{float}
\usepackage{soul}
\sethlcolor{yellow}
\begin{document}

\renewcommand{\hl}[1]{#1}


%\title[Assessing Control with Smart Home Cameras]{"I am Confident with what I do by hand, Automation can Break Down or Fail": Assessing Control with Smart Home Cameras}

\title[DataCrumb: A Physical Probe for Reflections on Background Web Tracking]{DataCrumb: A Physical Probe for Reflections on Background Web Tracking}


\author{Sujay Shalawadi}
\email{sujay.shalawadi@ntnu.no}
\orcid{0000-0003-3937-5427}
\affiliation{%
  \institution{Norwegian University of Science and Technology}
  %\streetaddress{Raufossvegen 40}
  %\postcode{2821}
  \city{Gjøvik}
  \country{Norway}
}

\author{Katrina Hvítklett}
\email{katrina.hvitklett@hotmail.com}
\orcid{0009-0001-0804-3887}
\affiliation{%
  \institution{Aalborg University}
   \city{Aalborg}
  \country{Denmark}}

\author{Anna Stentoft Ries}
\email{annastentoft@gmail.com}
\orcid{0009-0005-3332-1251}
\affiliation{%
  \institution{Aalborg University}
   \city{Aalborg}
  \country{Denmark}}

\author{Aisho Mohamed Ali}
\email{aisho-mohamed@outlook.com}
\orcid{0009-0001-9052-7185}
\affiliation{%
  \institution{Aalborg University}
   \city{Aalborg}
  \country{Denmark}}

\author{Florian Echtler}
\authornote{Corresponding author.}
\email{floech@cs.aau.dk}
\orcid{0000-0002-7175-9503}
\affiliation{%
  \institution{Aalborg University}
    \city{Aalborg}
  \country{Denmark}}

\renewcommand{\shortauthors}{Shalawadi et al.}

\begin{abstract}
%%129 words %%
Cookie banners and privacy settings attempt to give users a sense of control over how their personal data is collected and used, but background tracking of personal information often continues unnoticed. To explore how such invisible data collection might be made more perceptible, we present DataCrumb, a physical probe that reacts in real-time to data tracking with visual and auditory feedback. Using a research-through-design approach, we deployed the artifact in three households and studied participants' responses. Instead of providing details about what data was being tracked, the artifact introduced subtle disruptions that made background data flows harder to ignore. Participants described new forms of awareness, contradiction, and fatigue. Our findings show how sensory feedback can support reflection by drawing attention to tracking data flows that are usually hidden. We argue for designing systems that foster awareness and interpretation, especially when the users' control and understanding are limited.
\end{abstract}

\begin{CCSXML}
<ccs2012>
   <concept>
       <concept_id>10003120.10003138.10011767</concept_id>
       <concept_desc>Human-centered computing~Empirical studies in ubiquitous and mobile computing</concept_desc>
       <concept_significance>500</concept_significance>
       </concept>
   <concept>
       <concept_id>10003120.10003121.10003129</concept_id>
       <concept_desc>Human-centered computing~Interactive systems and tools</concept_desc>
       <concept_significance>300</concept_significance>
       </concept>
 </ccs2012>
\end{CCSXML}

\ccsdesc[500]{Human-centered computing~Empirical studies in ubiquitous and mobile computing}
\ccsdesc[300]{Human-centered computing~Interactive systems and tools}

\keywords{Physical Artifact, Reflective Design, Web Tracking, Privacy}

\begin{teaserfigure}
  \centering
  \includegraphics[width=0.475\textwidth]{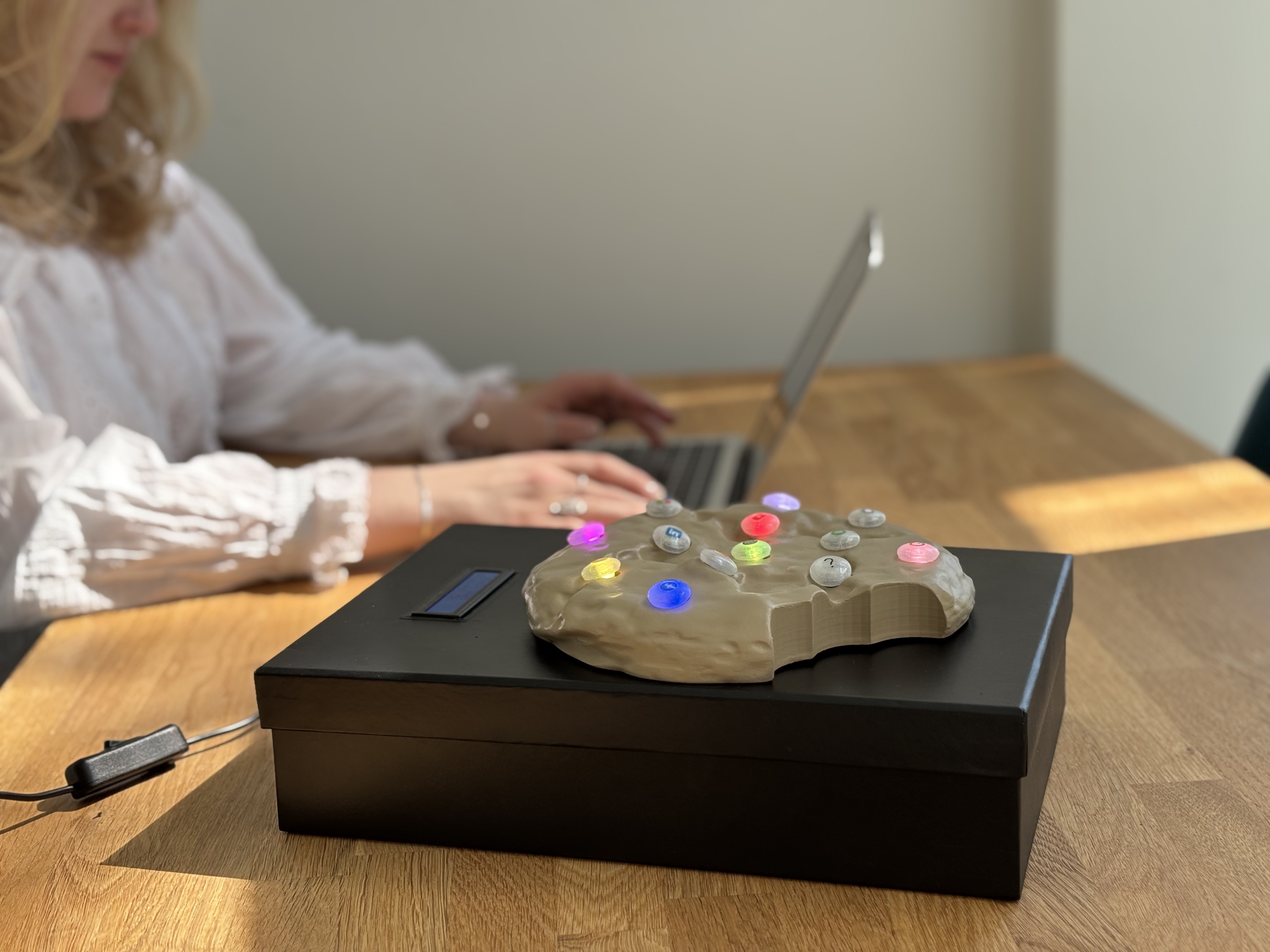}
  \hspace{0.05\textwidth}
  \includegraphics[width=0.46\textwidth]{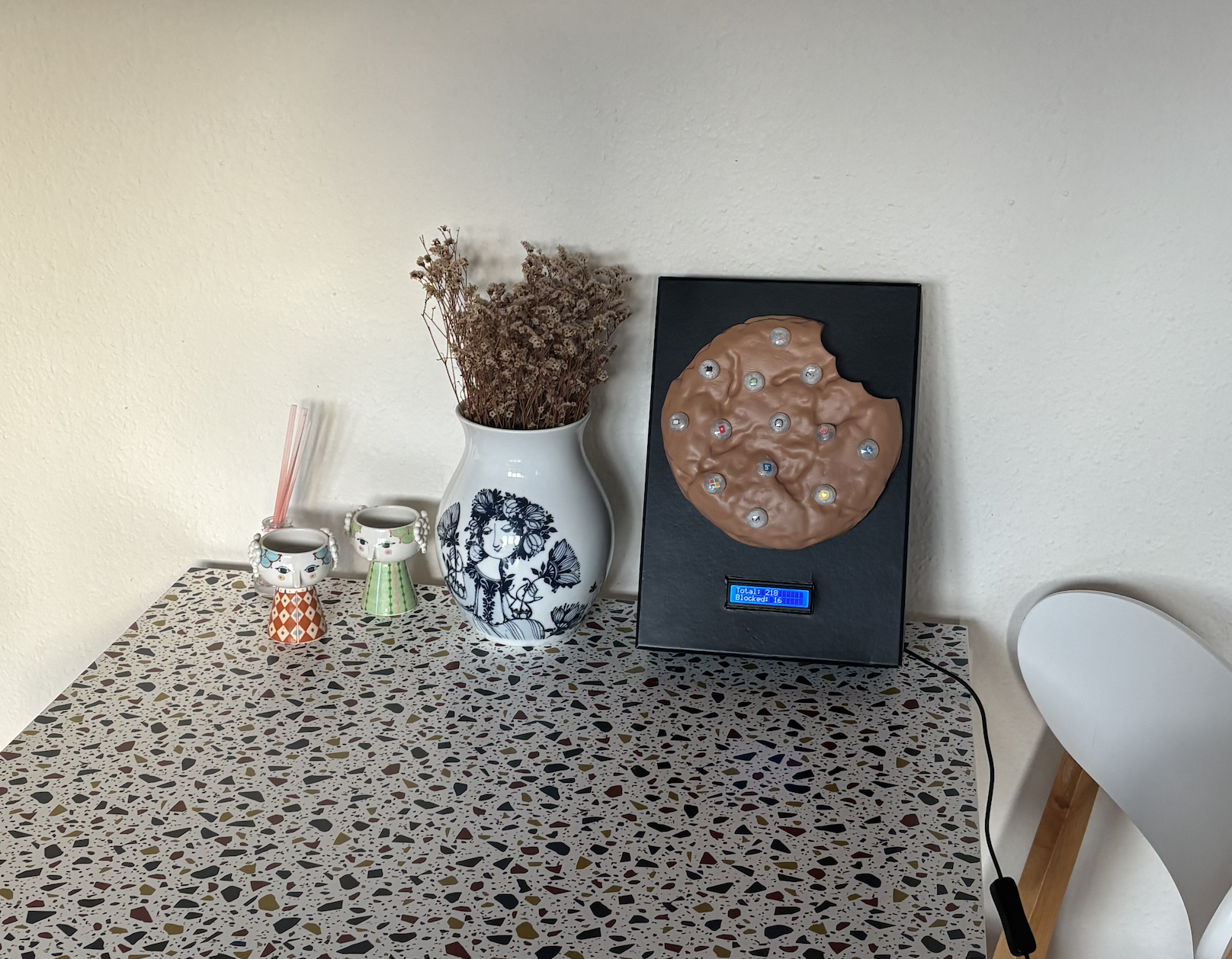}
  \caption{DataCrumb in use, in office (left) and domestic (right) environments.}
  \Description{A wooden desk with a big plastic cookie with LEDs on top in the foreground, and a woman working on a laptop in the background. Next to it, a closeup of the cookie with tech company logos on the LEDs, lit up in various colours.}
  \label{fig:teaser}
\end{teaserfigure}

\maketitle

\section{Introduction}

Online privacy controls such as cookie banners and consent interfaces have become routine due to legal requirements like the EU’s General Data Protection Regulation (GDPR), which mandates informed consent~\cite{DarkPatterns,CookieNotices}. Intended to inform users and offer choice, these standardized prompts, with familiar layouts and defaults, are often dismissed out of habit~\cite{CookieConsent}, leading to fatigue~\cite{Privacyfatigue} and disengagement~\cite{burkhardt2023privacy}. Even when users reject tracking, there is rarely visible feedback about what changes, making it difficult to know whether privacy is actually protected~\cite{DarkPatterns}.  

Despite efforts to reject cookies, adjust browser settings, or use ad blockers, background data collection often continues through third-party requests, fingerprinting, or DNS-level tracking~\cite{MURMANN2021101480,Papadogiannakis,rasaii2025intractable}. Operating quietly and without visible traces, these mechanisms bypass consent decisions and offer users few cues to sense when tracking occurs or whether their actions have any effect.  

The disconnect between perceived control and ongoing tracking illustrates how privacy is not simply a set of choices but an affective experience shaped by uncertainty and fatigue~\cite{Coopamootoo}. Technical protections may reduce exposure but offer little perceptible sense of what is happening in the moment. Prior work has shown that reflection can emerge not only from explanation or control but from noticing, contradiction, and disruption in everyday routines~\cite{Box,Rikke,VoxMox,FacebookDataShield}. This opens space to explore how physical artifacts might help users sense and interpret the ongoing presence of tracking, not to resolve it fully but to make it harder to ignore.  

We present \textit{DataCrumb}, a physical probe that responds to live tracking activity through light, sound, and a small screen (cf. figure~\ref{fig:teaser}). By translating background tracking into sensory feedback, DataCrumb invites users to reflect on whether their sense of privacy matches what is actually happening. Using a research-through-design (RtD) approach~\cite{RtD,RtD2}, we deployed the artifact in three households for three days each, complemented by pre- and post-interviews exploring how it supported reflection.  

Our contributions are threefold:
\begin{enumerate}
    \item We introduce DataCrumb, a physical artifact that transforms web tracking into a sensory experience through light, sound, and screen feedback.
    \item We present a design approach that uses sensory ambiguity and persistent signals to prompt reflection over time.
    \item We show how the artifact challenged users’ belief that tracking had stopped, revealing gaps between perceived and actual privacy.
\end{enumerate}

The rest of the paper reviews related work, describes the design and deployment of DataCrumb, presents three themes from its in-home use, and concludes with design implications and future directions.
\section{Related Work}

Relevant to our research are prior works in three areas: (1) the challenges of digital consent and the effects of privacy fatigue to maintain one's desired levels of privacy, (2) the role of materiality and tangibility in bringing awareness of digital data, and (3) the effects of provocation on dealing with ambiguous data tracking mechanisms.  

\subsection{Challenges in Digital Privacy}

Cookie banners and consent interfaces are now routine tools for communicating data collection online~\cite{CookieConsent}, introduced by regulations such as the GDPR to promote transparency and choice. Yet they are often dismissed due to habituation, limited understanding, and interface patterns that nudge acceptance~\cite{Nouwens,Utz}, leading to decision fatigue as users repeatedly manage complex choices.  

This gap between concern and behavior, often described as the privacy paradox~\cite{GERBER2018226}, reflects the burden of navigating opaque and manipulative systems rather than user irrationality~\cite{DarkPatterns}. Interface research has sought to clarify options~\cite{CookieDisclaimers} and strengthen user agency~\cite{CookieNotices}, but still treats privacy as a rational, momentary act~\cite{acquisti2015privacy,UbiCompConsent}. Fewer studies address how invisible data collection shapes users' emotions and sense of control over time~\cite{EmpoweringResignation}.  

In response, design research has explored how privacy can be experienced through material and sensory forms~\cite{VoxMox,InnerEar,Rogers,rosa2022frankie} showing that sensing and disruption rather than control can prompt reflection on hidden tracking. Building on this trajectory, our research examines how a tangible artifact can materialize cookie-based tracking in everyday contexts, not to restore control, but to expose the felt contradictions between perceived and actual privacy.

\subsection{Materializing Data Tracking through Tangible Interventions}

Standard online privacy interfaces such as cookie banners and consent forms have long struggled to render background tracking perceptible. While these systems meet legal requirements, they rarely evoke the lived implications of data flows in everyday contexts~\cite{Koelle}.  

In response, design researchers have explored material and sensory interventions that expose hidden digital processes through embodied interaction. \textit{VoxMox} disrupted smart-speaker use by combining live audio cues with a real-time listening timer, provoking reflection through mild discomfort~\cite{VoxMox}. \textit{Project Alias}~\cite{Alias} and the \textit{Privacy Hat}~\cite{PrivacyHat} used physical signaling and interference to resist voice-assistant surveillance. Large-scale installations such as the Facebook Data Shield visualized data sharing as a tangible interface between body and platform~\cite{FacebookDataShield}, while \textit{DataSlip} transformed invisible data transfers into printed slips that made collection visible and emotionally resonant~\cite{DataSlip}.  

Complementary work by Mehta et al. used image-schema meta\-phors~\cite{Mehta} and card-based toolkits~\cite{Mehta1} to support physical reasoning and hybrid interactions. Recent extensions explore algorithmic and sensor-based systems: Tangible LLMs gives physical form to machine learning models to foster interpretability~\cite{TangibleLLMs}, and SensorBricks uses modular play to build data literacy and collaborative understanding of sensing~\cite{SensorBricks}. Beyond privacy, The Troubling Cups highlights how material provocation and discomfort sustain engagement with structural inequalities~\cite{Naja}.  

Together, the above work show how emotional and embodied engagement rather than awareness alone can shape how people relate to hidden digital systems. They challenge the notion of privacy as purely rational control, revealing how confusion and disruption can make data activity perceptible. Building on this trajectory, our study explores how multimodal, real-time sensory feedback can materialize cookie-based tracking in everyday contexts, using persistence and ambiguity to prompt reflection rather than compliance.

\begin{figure*}[ht]
\centering
\includegraphics[width=0.8\linewidth]{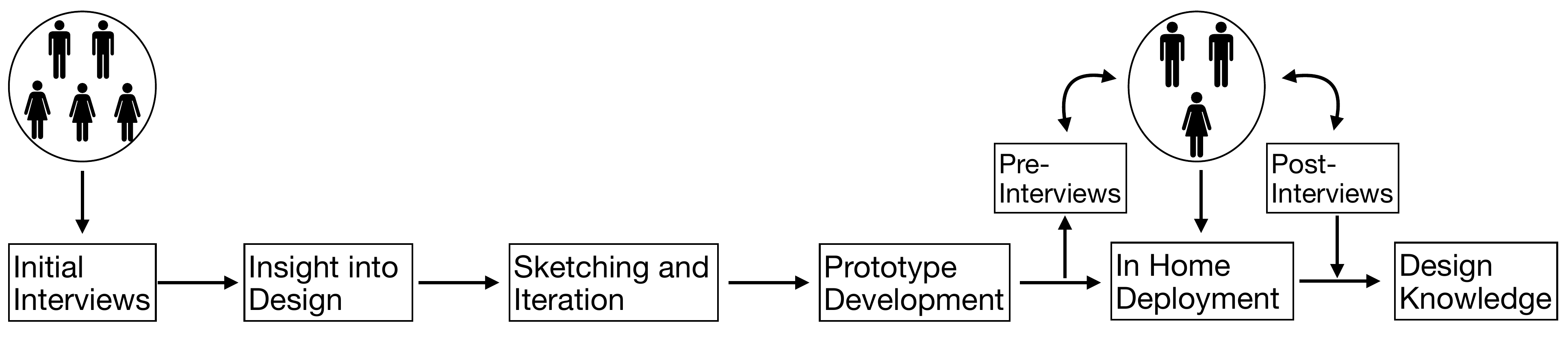}
\caption{Overview of our Research through Design process, from initial inquiry to deployment and reflection.}
\Description{Overview of our Research through Design process, from initial inquiry to deployment and reflection.}
\label{fig:rtdflow}
\end{figure*}

\subsection{Provocation and Reflective Design}

Design research often supports reflection through experience rather than explanation or instruction~\cite{Box}. Approaches that rely on physical presence, disruption, or repetition help people notice aspects of systems that are usually hidden or ignored~\cite{RevistingReflection,TangibleAffect}. Our work extends this orientation to web tracking by introducing a sensory artifact that makes background data flows harder to ignore. Instead of guiding privacy decisions, the prototype foregrounds tracking through repeated interruptions, prompting reflection as a lived and embodied encounter rather than a matter of control.  

Recent work expands the role of provocation in tangible interaction, showing how it can surface contested values and challenge normative assumptions. Jensen et al.~\cite{Rikke} show how provocation can transform functional objects into things for interpretation. Sensing Bodies~\cite{SensingBodies} uses material tension to reveal sociopolitical entanglements, while Toebosch et al.~\cite{SensorPI} propose open-ended sensor designs that invite personal interpretation. Together, this body of work demonstrate how ambiguity and discomfort can make hidden systems perceptible—an approach we adopt to explore reflection on background web tracking.

\section{DataCrumb: Design Intentions and Rationale}

\textit{DataCrumb} is a physical artifact designed to make online tracking more perceptible in everyday life. While many users take steps to block cookies or enable browser protections, these actions often offer little feedback about whether tracking continues. We did not aim to create a tool for control or explanation, but a physical presence that could bring background tracking into awareness through sensory experience.

\subsection{Research through Design Approach}

We adopted a Research through Design (RtD) approach~\cite{RtD,RtD2} to explore how a physical representation of cookie tracking might support everyday reflection on digital privacy. Our goal was not to solve a predefined problem or optimize a product. Instead, we used the artifact to provoke engagement and make hidden data activity more tangible in daily life.

Our approach builds on Zimmerman et al.'s~\cite{RtD} framing of RtD as a way to generate knowledge through iterative design, and Gaver's~\cite{RtD2} emphasis on raising new questions rather than converging on clear solutions. We treated DataCrumb not as a finished solution but as a deliberately interpretive artifact. Its function was to reveal how users notice, interpret, or live with tracking when it is made perceptible through physical and sensory feedback.

Figure~\ref{fig:rtdflow} outlines the four main stages of our design inquiry.

We began with a set of initial semi-structured interviews (see \autoref{Interviews}) to understand how users engage with cookie banners and tracking interfaces in everyday life. These conversations confirmed that while participants were aware of tracking, they often felt uncertain or fatigued by repeated consent requests. This informed our insight into design: rather than offering more settings or controls, we would design an artifact that reveals what users already try to manage but rarely perceive.

During the sketching and iteration phase, we explored different forms and feedback modalities, focusing on how a device might remain present without requiring interaction. The combination of light, sound, and display emerged from this phase as a way to support varied forms of attention — ambient, disruptive, and cumulative.

%We developed a working prototype using a Raspberry Pi and Pi-hole for DNS-level monitoring. In the in-home deployment, DataCrumb was installed in three households for three days each. It responded to network activity through light, sound, and display, without requiring user interaction or input.

Finally, our qualitative analysis focused on how participants experienced and interpreted the artifact over time. As detailed later in ~\autoref{Analysis}, this led to three themes that describe different forms of reflection supported by the probe.

\begin{figure*}[ht]
\centering
\begin{subfigure}[b]{0.28\linewidth}
\includegraphics[width=\linewidth]{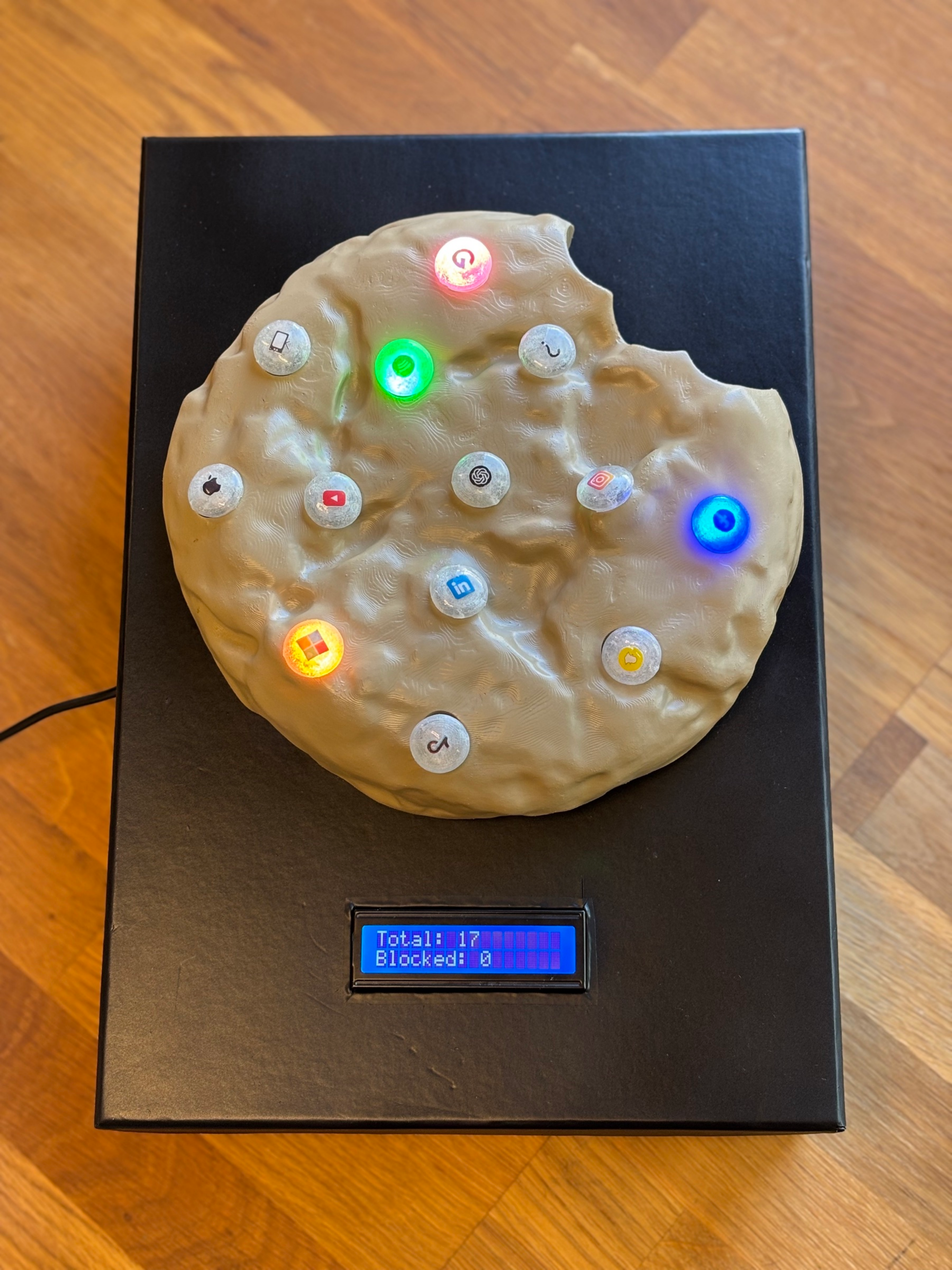}
\caption{Top view of DataCrumb showing LED layout and screen.}
\label{fig:topview}
\end{subfigure}
\hfill
\begin{subfigure}[b]{0.68\linewidth}
\includegraphics[width=\linewidth]{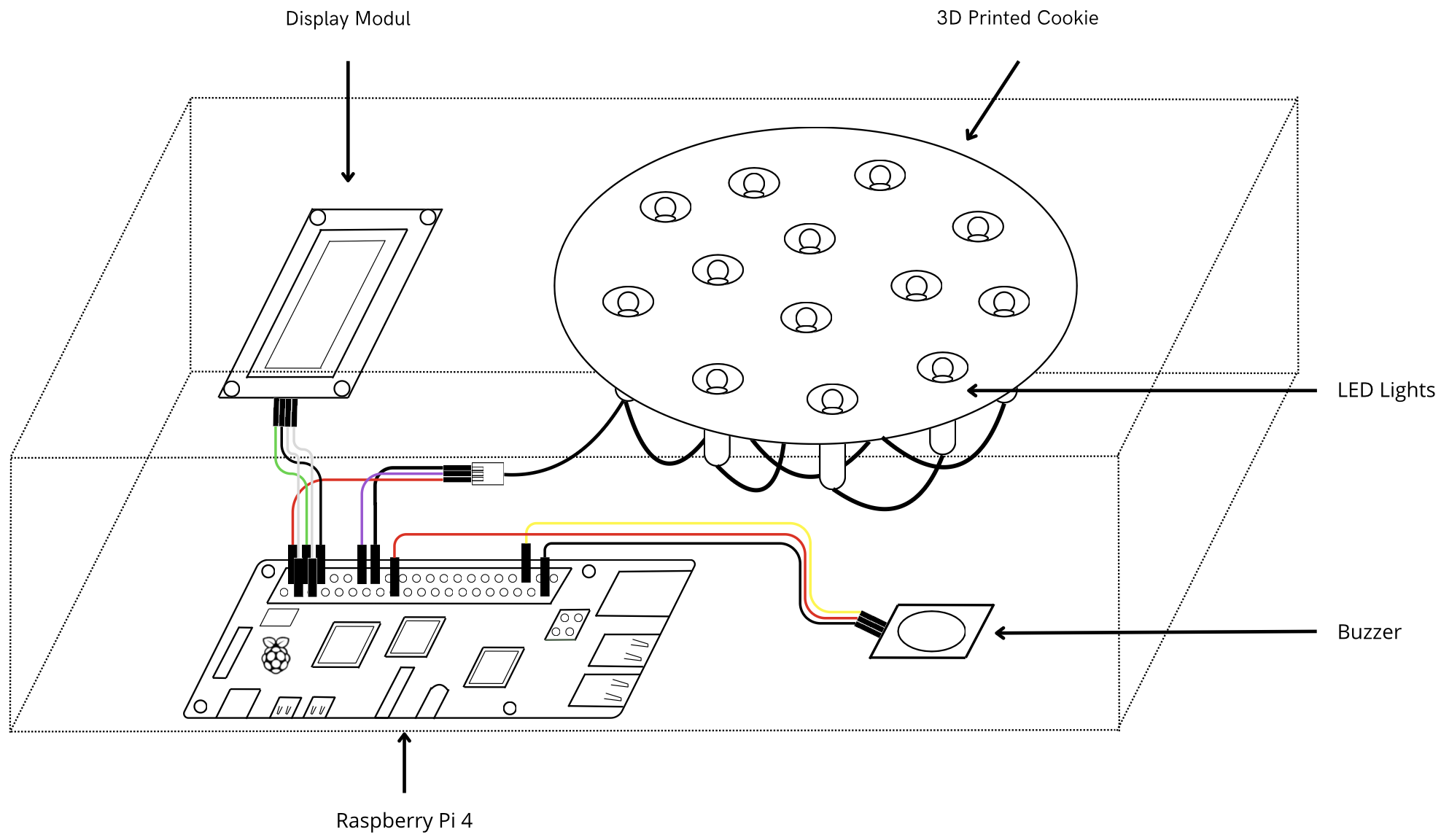}
\caption{Schematic view of DataCrumb components.}
\label{fig:insitu}
\end{subfigure}
\caption{The DataCrumb prototype. Left: artifact design and outputs. Right: internal components.}
\label{fig:artifactviews}
\Description{The DataCrumb prototype. Left: closeup of the plastic cookie, with LED lights labeled with tech company logos and lit up in bright colors. Right: schematic of internal components.}
\end{figure*}

\subsection{Artifact Design}

The artifact draws on the metaphor of the browser cookie, reimagining it as a persistent object in domestic space. This choice was both symbolic and functional: the familiar form based on the widely used "cookie" icon grounded engagement, while its behavior introduced unfamiliar signals that prompted noticing and interpretation. Rather than guiding users toward privacy choices, \textit{DataCrumb} was designed to sit within routines and occasionally disrupt them. 

Inspired by prior work that uses sensory cues (e.g., physicalization of data tracking) to materialize invisible data flows~\cite{DataSlip,VoxMox}, we designed the artifact to operate through light, sound, and a small display. These outputs are intentionally ambiguous. This ambiguity was designed to prompt interpretation rather than provide clear feedback. We wanted to encourage participants to question their privacy assumptions over time. Similar design strategies have been used in prior work where sensory cues are used to draw attention without fully explaining the underlying data collection, encouraging users to reflect on invisible digital processes~\cite{VoxMox,DataSlip,FacebookDataShield}. In our artifact, the LEDs signal blocked requests to known tracking domains, the buzzer emits short tones when blocking occurs, and the display shows cumulative activity of both permitted and blocked requests. These cues support gradual, situated awareness by encouraging users to interpret them over time, aligning with Bentvelzen et al.'s~\cite{RevistingReflection} view that reflection emerges through open-ended engagement rather than direct explanation.

The use of repetition, low-level disruption, and ambient signaling builds on ideas explored in systems like SensorBricks~\cite{SensorBricks}, but shifts the focus from education to reflection. Our goal was not to teach users how tracking works, but to help them sense its presence as part of their everyday environment. This required a form that blended into the home while remaining visible enough to prompt interpretation when triggered.

\subsection{Output Modalities}

DataCrumb responds to browsing activity using three output channels:

\textbf{LED Lights.} Thirteen LEDs are embedded beneath translucent domes labeled with well-known platform icons (e.g., Google, TikTok, LinkedIn, ...). When a request to a corresponding tracking domain is detected and blocked, the related light activates for 20 seconds. This delay helps make repeated activity more noticeable. The layout supports peripheral awareness in ambient conditions.

\textbf{Buzzer.} The buzzer emits a short tone when a domain query is blocked and a longer tone when multiple blocks occur in succession. This cue was intended to create a momentary break in the environment which is audible enough to be noticed, but not detailed enough in order to encourage self-interpretation without explicit explanation.

\textbf{Display Screen.} A small digital display shows a live count of total domain requests and the number of blocks. This ongoing update created a slow, cumulative awareness of traffic volumes, even when no device was in active use.

\subsection{Technical Setup}

Our prototype was built using a Raspberry Pi 4b configured to run Pi-hole\cite{pihole}, a local proxy that monitors all requests to the Domain Name System (DNS) on the participants' home network and blocks known tracking or data collection requests. A custom Python script parses DNS logs in realtime, matches requests to known services (e.g., Facebook, Apple, OpenAI, ...), and triggers light and sound output through GPIO pins.

The system was designed to function autonomously throughout the three-day deployment, requiring no user input. This setup allowed the artifact to act as a passive presence that surfaced invisible tracking flows through ambient feedback rather than direct interaction.

%In the next section, we describe how this concept was developed and refined through a Research through Design (RtD) process and how early user insights informed design decisions.

\section{Evaluation}

Our research involved evaluation with external participants at two key stages. The first stage focused on understanding how people perceive and interact with cookie banners in everyday digital life. These insights helped shape the problem framing and initial concept development. The second stage involved deploying the DataCrumb prototype into people's homes to explore how users experienced it as part of their daily routines.

\subsection{Initial User Inquiry}
\label{Interviews}
The first phase of our study aimed to explore how people perceive and respond to cookie banners and tracking in everyday digital contexts. We conducted five semi-structured interviews with participants aged 25 to 33 (3 female, 2 male), all regular users of digital platforms but without specialized training in privacy technologies or system-level tools. They approached privacy from a user perspective, relying primarily on browser settings, built-in options, or default behaviors.

Interviews lasted 30 to 45 minutes and focused on participants' habits around cookie consent, awareness of tracking, and emotional responses to repeated privacy prompts. Questions included: ``What do you usually do when you see a cookie banner?'', ``Have you ever adjusted cookie settings?'', and ``How do you feel about being tracked online?'' Follow-up prompts encouraged participants to reflect on specific moments or changes in behavior over time.

We used affinity diagramming to analyze the transcripts. Statements were grouped around common patterns, then refined into two key insights that shaped our design direction:

\begin{enumerate}
    \item Perceived awareness, limited visibility: Participants believed they were taking steps to protect privacy—such as rejecting cookies or using ad blockers—but acknowledged that they had little visibility into whether tracking was actually stopped. This disconnect suggested a perceptual gap between action and outcome.
    
    \item Resignation and fatigue: While participants expressed concern, they also described feeling worn down by repeated prompts and unclear consequences. This aligns with concepts such as privacy fatigue~\cite{Whyprivacy,Privacyfatigue} and the privacy paradox~\cite{solove2021myth}, where stated concerns do not translate into meaningful action.
\end{enumerate}

These interviews confirmed our assumption that the challenge was not a lack of awareness or control options, but the absence of a way to perceive what tracking looks and feels like in practice. This insight helped frame our design intervention: rather than creating another consent tool, we would explore how a physical artifact could surface the presence of tracking and prompt reflection through lived experience. \hl{The analysis of these preliminary interviews followed a reflexive thematic orientation}~\cite{braunclarke2023reflexivethematic} \hl{consistent with our later deployment study. Two researchers independently reviewed the transcripts, highlighting recurring expressions and metaphors around tracking, control, and fatigue. Their observations were discussed in two analysis meetings over one week, during which preliminary codes were clustered into broader patterns. Through iterative memoing and comparison, these were refined into two higher-level insights: perceived awareness with limited visibility, and resignation and fatigue; which shaped the subsequent design phase and guided the development of DataCrumb.}

\subsection{In-Home Deployment}

After developing the DataCrumb prototype, we deployed it in three student households to examine how its presence might shape participants' perception of background tracking in everyday settings. Each deployment lasted three days and followed a consistent protocol: a short pre-interview, three days of autonomous artifact operation, and a post-interview focused on participants’ interpretations and experiences.

All three participants were students living in small domestic spaces. Participant A (25, male) lived with a roommate and placed the artifact in his bedroom. Participant B (25, female) lived alone and described herself as generally disinterested in technical configuration. Participant C (23, male) also lived alone and used no additional privacy tools beyond default browser settings. While their levels of digital literacy varied, all three participants used common devices and engaged with online services regularly. The focus was not on expertise, but on how everyday users might interpret tracking when it became materially present.

\hl{Participants for the in-home study were recruited through university mailing lists and word of mouth. We aimed for everyday users familiar with online platforms but without technical expertise in privacy tools. Participation was voluntary, and no compensation was offered. Each deployment followed the same three-day protocol, preceded and followed by semi-structured interviews. The data were reviewed collaboratively by two researchers using reflexive thematic analysis, consistent with the approach described in Section}~\ref{Analysis}.

The prototype was placed in a visible area of the home, such as a desk or bedside table, and connected to the home Wi-Fi network. Pi-hole ran in the background to intercept DNS traffic, triggering DataCrumb's outputs—LEDs, sound, and display—in real time. No interaction or configuration was required during the deployment. Participants were told the artifact would respond to data flows on their network but were not informed of the exact mappings between signals and tracking services. This allowed us to observe how meaning was constructed through use, not explanation.

The post-interview focused on what participants noticed, how they interpreted the signals, and whether the artifact prompted new thoughts or routines. We asked questions like ``When did you notice the artifact most?'', ``Did it affect your device use?'', and ``How did it make you think about tracking?'' These responses formed the basis for our thematic analysis, presented in the following sections. 

\hl{Across the three household deployments, DataCrumb operated continuously for approximately three days per site and logged between eight and fifteen activation events per day. Each activation corresponded to background data exchanges rather than deliberate browsing activity, producing a total of about twenty-five to forty notifications per household. These figures, derived from device logs and researcher field notes, help contextualize participants' later descriptions of ``constant'' or ``repetitive'' alerts. While quantification was not the focus of this RtD study, these observations situate the qualitative findings on attention, desensitization, and privacy fatigue discussed in the next sections.}

Because the artifact's signals could potentially provoke irritation, confusion, or concern, we took steps to ensure participants' comfort and autonomy. Each participant was informed that the artifact was experimental, did not collect personal or identifying data, and could be unplugged at any time without penalty. We clarified that they were under no obligation to keep it running and that their participation could be paused or withdrawn at any moment. We also explained that the artifact was intentionally ambiguous and that the goal was not to test them, but to learn how they experienced its presence.

\subsection{Analysis Process} \label{Analysis}

We used a reflexive thematic analysis approach~\cite{braunclarke2023reflexivethematic} to examine how participants experienced and interpreted the presence of DataCrumb in their everyday routines. Rather than coding for frequency or aiming for saturation, our goal was to construct situated interpretations of how reflection emerged over time in response to the artifact's ongoing presence.

Transcripts and notes from pre- and post-interviews were reviewed collaboratively. The second author, who conducted the deployments, shared memos highlighting moments of emotional tension, interpretive uncertainty, and shifting assumptions. These were then discussed with the co-authors to develop alternative readings and extend the analytical framing. This collaborative approach allowed us to remain attentive to both situated detail and broader experiential patterns.

We organized key quotes into provisional clusters in a shared document. Initial groupings such as ``confidence in control'', ``unseen activity'', and ``buzzer reactions'' served as entry points for interpretation. Through iterative discussion, these clusters developed into the three themes that structure our findings. For example, early confidence in privacy settings gave way to doubt when tracking persisted, forming the basis for \emph{From Control to Contradiction}. Repeated feedback gradually became noticeable, supporting the theme \emph{Sensing the Invisible}. What began as annoyance toward the buzzer was later interpreted as a prompt for reflection, shaping \emph{Disruption as Reflection}.

Each theme follows a temporal structure—before, during, and after deployment—to trace how participants’ attention, confidence, and interpretations shifted in response to the artifact’s signals. The themes do not aim to generalize across users, but to surface how reflective engagement unfolded through lived interaction with the prototype.

\section{Findings} \label{Findings}

\hl{The following three themes explore how participants responded to the presence of DataCrumb in their homes, each tracing a distinct form of engagement with background data tracking. Rather than presenting overlapping observations, the themes address different dimensions of the experience: cognitive shifts in perceived control, sensory awareness of hidden activity, and affective responses to disruption. The first theme captures how participants' confidence in managing privacy gave way to uncertainty and contradiction. The second shows how repeated sensory cues transformed abstract knowledge into situated awareness. The third highlights how discomfort and disruption, particularly through sound, prompted reflection by interrupting routine. Together, these themes illustrate how a tangible artifact can challenge assumptions, reorient attention, and provoke reflection through its ongoing presence.}

\subsection{From Control to Contradiction}
This theme traces how participants' confidence in managing their digital privacy shifted over the course of the study. In pre-interviews, participants expressed a strong sense of agency and routine in navigating cookie banners. During the deployment, DataCrumb’s real-time responses disrupted those assumptions. In post-interviews, participants reflected on the mismatch between their intentions and the continuous background activity they observed.

\paragraph{Pre: Privacy as Routine and Under Control.}In the pre-interviews, all three participants described themselves as cautious and deliberate in managing cookie tracking. P1 stated, \textit{``I always reject cookies. I’ve set up my browser to block most of them.''} P2 shared a similar view: \textit{``I try to limit what’s shared. I always go in and choose the minimum.''} P3 also emphasized a sense of control, explaining, \textit{``I use Brave because it blocks stuff automatically. I don’t trust websites to be honest about what they track.''}

Participants treated these steps as a kind of set-it-and-forget-it approach to privacy. Their routines gave them confidence, but they were rarely questioned. As P1 put it, \textit{``I think it’s more habit at this point. I know what to click. I don’t even think about it now.''} These actions had become automatic, offering a sense of control without much reflection.

This reliance on routine became visible only when the artifact introduced unexpected signals that disrupted that familiarity. For P2, one such moment came when the prototype responded unexpectedly: \textit{``I think I was just sitting and not even using the laptop, and it buzzed. That kind of shocked me.''} These moments of disruption created space for participants to reconsider practices they had previously taken for granted.

\paragraph{Mid: Disruption of Assumptions Through Real-Time Feedback.}During the deployment, the prototype’s sensory responses often activated at unexpected times, especially when participants were not actively using a browser or app. This caused confusion and raised doubts about what had actually been agreed to or blocked. Rather than confirming a gap that already existed, the artifact actively disrupted participants' confidence in their prior consent decisions by revealing background tracking they had not anticipated.

These disruptions prompted second-guessing. P2 shared, \textit{``Sometimes I wondered, is it doing that because of something I didn’t say no to? Like, maybe I missed a banner.''} Although participants had previously expressed confidence in their choices, these unexpected signals forced them to reflect on what might have slipped through. The prototype didn’t provide explanations; it simply surfaced moments of data traffic in real time. But that lack of clarity was part of what made it unsettling. As P1 noted, \textit{``There were times it lit up and I was like… I didn’t open anything. That was kind of worrying.''}

This worry did not translate into major changes in behavior, but it created friction that lingered. Rather than correcting or resolving participants’ concerns, the prototype left them uncertain. As P3 put it, \textit{``It caught stuff I wouldn’t have noticed. I guess that’s the point.''} These moments of noticing were not always actionable, but they exposed how tracking mechanisms often operate in the background, without clear feedback or limits. The prototype made it clear that rejecting cookies did not prevent all tracking, and it exposed how much background data traffic continued without participants’ knowledge or control.

\paragraph{Post: Re-evaluating Confidence and Encountering Fatigue.} \hl{Across the three households, DataCrumb activated roughly eight to fifteen times per day, often when participants were not directly interacting with their devices. This rhythm of alerts were small but persistent, shaped how fatigue accumulated over time.}. 

By the end of the deployment, participants began to question the confidence they had expressed earlier. Their belief in being careful and in control gave way to more conflicted reflections. They still valued privacy, but started to acknowledge that their actions did not always align with those values. P2 admitted, \textit{``Sometimes I just want that damn banner to go away, so I accept all.''} P1 similarly noted, \textit{``There are some sites I use every day where I’ve probably just clicked accept.''} These admissions were not framed as careless mistakes, but as compromises shaped by repetition, fatigue, and convenience.

Alongside this shift, participants described a growing sense of emotional fatigue. P2 explained, \textit{``It feels like the system is designed to make you tired.''} P3 added, \textit{``You feel like you’ve done your part, but then it keeps lighting up. You just get tired.''} The fatigue was not due to confusion, but from seeing that their actions of rejecting cookies or adjusting settings did not seem to reduce the background activity as they had always imagined.

These responses marked a turning point. As P1 put it, \textit{``It kind of made me second guess whether I was as careful as I thought.''} While the artifact didn’t teach participants anything new, it reframed what they thought they knew—turning assumptions into questions. It created moments where participants could reflect on what they believed they had already controlled, and where that control began to feel uncertain or incomplete.

\hl{Having shown how participants' confidence in managing privacy gave way to contradiction and fatigue, the next theme turns to how those realizations became perceptible in daily life. It explores how repeated sensory cues made tracking feel present rather than abstract.}

\subsection{Sensing the Invisible}

This theme examines how the prototype shifted participants' understanding of data tracking from an abstract concept to a tangible presence. While participants were generally aware that digital tracking occurred in the background, the artifact translated that invisible activity into sensory cues. Over time, the lights, sound, and display became part of the domestic environment, allowing participants to develop a new attentiveness to the timing, frequency, and scale of ongoing data exchanges. The theme traces this shift across the deployment, from abstract awareness to situated noticing.

\paragraph{Pre: Awareness Without Presence.}
Before the deployment, participants acknowledged that data tracking occurred in the background, but this awareness was abstract and rarely acted upon. P3 noted, \textit{``I know apps talk to each other or send stuff, but it’s not something I think about unless there’s a problem.''} P2 expressed a similar view: \textit{``I mean, I know websites collect stuff, but it’s not like I can see it. I just click what I have to and move on.''} These reflections suggest that participants understood the presence of tracking but did not perceive it as an active part of their environment.

For most, this awareness was triggered only by visible interruptions, such as pop-ups or banner prompts. As P1 explained, \textit{``Unless I see something pop up, I don’t really think about what’s going on in the background.''} These comments show that tracking was framed as distant or technical, not something embedded in the routines of daily life. The idea of background data activity was recognized, but it remained outside the boundaries of everyday attention or concern.

\paragraph{Mid: Noticing the Invisible Through Repetition.}  
During the deployment, the prototype introduced sensory responses that brought this background activity into focus. Participants began to notice how often data moved, even when they were not actively engaging with their devices. P2 described one such moment: \textit{``It lit up when I was just watching something on my laptop. Not even browsing. That felt a bit creepy.''} P1 recalled, \textit{``There were a few times it flashed and I had no idea why. That stuck with me.''}

As these signals repeated, participants began to interpret them as meaningful. P3 observed, \textit{``It felt like YouTube and TikTok were always triggering something. Even when I wasn’t directly using them.''} P2 noted, \textit{``You kind of start picking up on which ones are always lighting up. It’s not random.''} These patterns emerged gradually, not through instruction, but through repeated exposure to the device’s behavior.

The display screen added another layer of awareness. P1 explained, \textit{``The number going up felt strange. Like, you’re not doing anything, but something’s still happening.''} The accumulation of blocked and active requests created a sense of scale that had previously been missing. Participants described the feedback as low in intensity but consistently noticeable. It did not demand active monitoring, yet it remained perceptible within the flow of their everyday routines and gradually informed their sense of ongoing data activity. \hl{Each household's log showed a similar cadence of between eight and fifteen activations per day, meaning that repetition itself became part of the experience. Participants did not need to count events and the steady recurrence was enough to transform abstract awareness into lived texture for sensing the invisible.}

\paragraph{Post: Data Tracking as Embedded Presence.}  
By the end of the deployment, participants had developed a new sensitivity to the invisible activity around them. The prototype did not require interaction, but it became a part of the environment that quietly shaped their awareness. P2 described this shift: \textit{``It was always there, like a part of the room. You’d glance at it or hear a beep and just know something was going on.''} 

This shift was not dramatic, but it changed how participants understood their devices. P3 reflected, \textit{``It kind of changed the way I thought about my devices. Like, even when they’re not doing anything, they’re still talking to someone.''} P1 added, \textit{``You start seeing things differently. Even the quiet moments have something happening in them.''}

The artifact’s presence made background tracking perceptible, not through alerts or explanations, but through repeated sensory cues embedded in domestic space. What was once abstract and non-perceptible became something local and continuous that something, participants learned to notice in everyday life.

\subsection{Disruption as Reflection}

\hl{Building on this gradual awareness, the final theme examines how discomfort, especially through sound, pushed participants from passive noticing toward active reflection.} This theme explores how discomfort, rather than subtlety, can support reflection on background tracking. While the lights and display became part of the home environment, the buzzer introduced moments of irritation that broke through passive attention. These interruptions did not produce clarity or delight but triggered awareness through tension. The theme follows how participants’ initial expectations of ambient feedback were disrupted, leading to moments of interpretation and behavioral response.

\paragraph{Pre: Expecting Calm, Visual Feedback.}Before the deployment, participants assumed the artifact would offer passive, ambient feedback. They anticipated something that would stay in the background and remain unobtrusive. P2 noted, \textit{"I thought it would just blink or do something lowkey, not beep at me."} P1 shared a similar expectation: \textit{"I assumed it would just sit there and blink now and then."} P3 added, \textit{"I didn’t expect it to make sound. I thought it was more like a mood light or something."} These remarks reflected a shared belief that the prototype would be more aesthetic than affective, offering information quietly without disrupting routines.

\paragraph{Mid: Irritation as an Unexpected Signal.} Once deployed, the buzzer became the most noticed feature, not because it conveyed new information, but because it refused to be ignored. P2 stated, \textit{"It was annoying, but impossible to ignore."} P1 expressed their frustration, \textit{"I think I told it to shut up more than once."} Unlike the other outputs, the buzzer introduced auditory friction that punctuated ordinary moments and demanded attention. \hl{Across deployments, these audible cues occurred several times each day that was enough to interrupt routines without becoming constant. Participants described this frequency as tolerable yet tiring, a balance that sustained awareness while revealing how quickly attentiveness could turn into fatigue.}

This discomfort, while unpleasant, led to reflection. The sound often triggered questions and curiosity. P3 noted, \textit{"It made me curious—what was it blocking? Why now?"} Even though the prototype offered no explanation, its unpredictable responses pushed participants to wonder about the hidden data flows behind their everyday device use. P2 described the emotional effect of this pattern: \textit{"It was like the house was telling me I was doing something wrong."}

\paragraph{Post: Attentiveness Through Discomfort.}By the end of the study, participants still disliked the buzzer but began to understand its role. P3 described a small behavioral shift: \textit{"I actually put my phone away earlier just so it wouldn’t keep beeping."} This change was not dramatic, but it reflected how irritation could become a cue for pausing or reconsidering one's habits.

Participants did not see the buzzer as informative or actionable. It did not clarify what was being tracked or allow them to respond directly. However, they described it as something that interrupted their routines and drew attention to tracking in ways other signals did not. This disruption made tracking feel more present and unavoidable. For some, the discomfort triggered curiosity or skepticism about what their devices were doing. Rather than fading into the background, the sound prompted moments of reflection on whether their privacy settings were working as intended.

\section{Discussion}

In this section, we reflect on what our findings suggest for designing artifacts that support privacy reflection in everyday life. DataCrumb did not provide users with control or detailed explanations; instead, it used sensory signals to disrupt routines and reveal that tracking continued even when participants believed it had stopped. \hl{While earlier artifacts such as DataSlip}~\cite{DataSlip}, \hl{VoxMox}~\cite{VoxMox}, \hl{and Project Alias}~\cite{Alias} \hl{made hidden data flows visible, DataCrumb advances this work by revealing tracking as it happens through an everyday domestic artifact. By linking real-time network activity to light, sound, and display cues, it transforms privacy awareness into a sensory and affective experience rather than a purely informational one.} We structure the discussion around two main insights: how contradiction and disruption prompted reflection, and how repeated signals made tracking feel present, before concluding with design considerations and limitations of our approach.

\subsection{Reflection Through Contradiction and Disruption}

Participants began to reflect when the artifact signaled tracking activity even after they had rejected cookies, blocked ads, or used privacy browsers. As seen in theme 1, \textit{From Control to Contradiction}, these routines had become habitual markers of control. When DataCrumb responded to DNS-level activity despite these precautions, it quietly undermined this confidence. \hl{During the three-day deployment, the artifact signaled often enough to be noticed in everyday life without taking over. Over time, this steady pattern changed participants' reactions from curiosity to tiredness, showing the kind of fatigue that comes from constant reminders about privacy}~\cite{Privacyfatigue}. \hl{Repetition became part of the reflection itself, as ongoing awareness slowly turned into weariness.} Reflection emerged not from guidance but from the mismatch between expectation and experience, consistent with RtD approaches that surface questions rather than answers~\cite{RtD2}.  

This tension echoes Nissenbaum's notion of contextual integrity~\cite{nissenbaum2009privacy} where discomfort arises when established norms of consent no longer align with technical realities. Participants' consent routines that were shaped by repetitive banners that promote habitual clicking~\cite{Utz,Nouwens} were now disrupted as the artifact exposed the persistence of tracking. The design thereby interrupted a cycle of passive engagement and invited participants to confront the limits of their assumptions about privacy control.  

Comparable strategies appear in reflective design that employs indirect or ambiguous feedback~\cite{RevistingReflection}. DataCrumb's signals were embedded in everyday routines, encouraging observation rather than decision-making. The buzzer extended this effect by adding sensory friction: an irritation that momentarily stopped activity and made tracking impossible to ignore. Similar uses of discomfort appear in VoxMox, which used sound to reveal voice-assistant listening~\cite{VoxMox}; in DataSlip, which printed physical receipts of data exchange~\cite{DataSlip}; and in The Troubling Cups, which used unease to sustain attention on structural inequality~\cite{Naja}. These works, like ours, treat interruption as a resource for reflection rather than a means of resolution.  

Such provocation raises ethical questions about what designers ask users to endure. DataCrumb neither stopped tracking nor provided clarity, but made its persistence tangible, leaving participants to interpret their own responses. Although they were told they could unplug it at any time, all chose to keep it running suggesting that discomfort was voluntarily sustained as part of engagement. \hl{The challenge for designers is that reflective provocation always involves a degree of exposure: users are asked to confront unsettling realities without clear resolution. In our case, the repeated cues rendered tracking tangible, but also demanded emotional energy to stay attentive. This revealed a tension between the value of discomfort as a reflective trigger and the risk of exhausting participants in the process. Ethical reflection, therefore, is not only about what a design reveals, but about how long and how intensely it keeps users in that state of unease.}  

\hl{Although the deployment was brief, participants chose to keep the artifact running even when it became tiring. This short-term endurance suggests that discomfort, when bounded and voluntary, can support critical reflection. Yet it also underscores a design responsibility: prolonged exposure risks shifting from provocation to burden. Designers must therefore provide space for disengagement and recovery, ensuring that reflection through friction remains constructive rather than depleting.}  

In RtD, this involves giving participants room to step back or reinterpret the experience while attending to how strongly a design demands attention. DataCrumb's signals made tracking difficult to dismiss; handled with care, such friction can be as valuable for reflection as providing explicit solutions.

\subsection{Noticing Tracking and Feeling Its Weight}

Participants were already aware that online tracking happens in the background, often beyond their perception. But the artifact made this awareness more immediate. Through repeated cues from lights, sounds, and screen feedback, it gave physical form to the presence of data flows that participants associated with privacy-invasive tracking. These cues became linked to familiar routines such as opening an app, browsing late at night, or idling on a device. \hl{Across the three households, this activity occurred several times each day, enough to remain noticeable without dominating attention. Over time, the steady rhythm of signals transformed awareness from abstract knowledge into something physically and emotionally felt.} Although the feedback was persistent, it did not explain or interrupt. Rather than demanding attention all at once, the feedback gradually shaped how participants noticed when tracking occurred. As with prior work on systems designed to stimulate reflection~\cite{RevistingReflection,TangibleAffect}, the artifact's steady presence made the invisible slightly more perceptible. In this way, tracking shifted from an abstract concern to something continuously and bodily felt.

The artifact made background tracking perceptible, but did not explain what was being collected or by whom. Its signals were deliberately ambiguous. While participants occasionally linked blinking lights to specific platforms through close inspection, the feedback offered no confirmation beyond that. Similar to other tangible systems that expose hidden infrastructures~\cite{InnerEar,DataSlip}, the artifact did not resolve uncertainty but gave form to what participants had long suspected but rarely perceived. In response, participants returned to familiar actions such as rejecting more cookies and adjusting settings more frequently. These efforts, however, offered little relief. Over time, participants described feeling worn out by the repetition. This affective response aligns with prior work on privacy fatigue~\cite{Privacyfatigue,Whyprivacy}, not as a failure to act but as an emotional condition shaped by repeated exposure and limited control. This fatigue often led to resignation. As Seberger et al.~\cite{EmpoweringResignation} argue, resignation is not necessarily apathy, but a rational response to systems that offer few meaningful alternatives. Participants did not feel empowered, but they also did not disengage. They continued to pay attention, even when their actions felt ineffective. Reflection in this case was not sustained through agency, but through a growing awareness of its limits. DataCrumb did not guide participants toward specific insights. Instead, it became part of the home environment, remaining present without resolving. It supported reflection not by offering answers, but by allowing participants to feel what they could not fully explain, yet could no longer ignore.

\subsection{Design Implications}

Based on our findings, we identify four design implications relevant for designers of tangible interactive system interventions that aim to alleviate privacy concerns in relation to data tracking in domestic settings.

\textbf{1. Support reflection through presence, not explanation.}  
Participants reflected not because the artifact explained what was happening, but because it made background tracking perceptible in subtle and repeated ways. Similar to prior work on ambiguity and slow design~\cite{RevistingReflection,Gaver}, the artifact created interpretive space rather than instructional guidance. Designers might consider how persistent, low-effort feedback can support reflection by allowing users to revisit their assumptions over time, without requiring explicit prompts or explanations.

\textbf{2. Use disruption carefully to prompt re-evaluation.}  
The buzzer was disruptive but not dismissed. Its relevance, rather than its intensity, made it meaningful in context. Prior work has shown how friction and discomfort can be used to surface hidden infrastructures and provoke critical attention~\cite{VoxMox,DataSlip,Naja}. Designers should not aim to eliminate discomfort entirely but might explore ways to introduce disruption in measured, context-aware forms that invite reflection without overwhelming users.

\textbf{3. Acknowledge fatigue as a meaningful response, not a failure.}  
\hl{Participants described privacy fatigue not as apathy but as a sensible reaction to continuous effort with limited visible effect. This aligns with research on emotional exhaustion in privacy management}~\cite{Privacyfatigue,Whyprivacy} \hl{and perspectives that frame resignation as a rational response to constrained agency}~\cite{EmpoweringResignation}. \hl{Rather than treating fatigue as a design flaw, it can be recognised as part of the reflective trajectory—an affective signal that users have reached the limits of agency. Designers could make this effort visible and valid, for example by showing when repeated rejections occur or by acknowledging protective actions that otherwise remain unseen. Such cues can affirm the user's ongoing engagement instead of concealing it.}

\textbf{4. Use repetition to shift background activity into awareness.}  
Reflection emerged gradually through repeated, low-intensity feedback. Participants noticed patterns over time, not through explanation, but through persistence. This supports prior work on ambient and tangible systems that reveal hidden processes through continued exposure~\cite{TangibleAffect,InnerEar}. We suggest that designers explore building small, recurring signals, such as the lights, sounds, or on-screen indicators used in our artifact, that can reveal background activity without actively disrupting users. Over time, such cues can help make processes related to tracking or data exchange more noticeable in everyday life, even if they remain technically opaque.

\subsection{Limitations and Future Directions}

This study involved a short-term deployment of a single artifact in three domestic contexts. While this aligns with RtD’s focus on situated interpretation, it limits generalizability; the insights are contextual rather than representative. Future work could explore how similar interventions unfold across varied household arrangements or privacy orientations.  

The prototype responded to DNS activity rather than cookie-level data, introducing ambiguity in what was being detected. Participants could not always link signals to specific behaviors, yet this uncertainty supported speculative engagement consistent with work that treats interpretive ambiguity as a design resource~\cite{Windl,Shalawadi}. Future iterations might experiment with optional scaffolding, as in PrivacyCube~\cite{PrivacyCube}, while preserving openness as a reflective quality.  

\hl{Future studies could extend this approach to longer deployments or larger participant groups, or pair qualitative reflection with lightweight quantitative measures such as usage logs or short surveys on perceived fatigue or awareness} (e.g.,~\cite{Rainmaker}). \hl{Such mixed-method extensions would preserve the interpretive depth of RtD while offering broader insight into how reflection sustains or fades over time.} Longer-term studies could also examine whether artifacts like DataCrumb remain reflective, fade into routine, or are reinterpreted over time~\cite{Windl}. Finally, the artifact’s outputs were fixed in form and intensity. While this constraint supported provocation, it limited adaptation. Future work could explore adjustable feedback without undermining the interpretive tension that ambiguity affords~\cite{Shalawadi}.

\section{Conclusion}

This paper presented DataCrumb, a physical probe that makes background web tracking more perceptible through audiovisual feedback. Rather than explaining or instructing, the artifact disrupted expectations and routines, prompting reflection through contradiction, irritation, and uncertainty. Participants began to notice the presence of tracking not as information, but as a lived experience that challenged their sense of control.

Our findings suggest that ambiguity, repetition, and discomfort can support situated reflection on data privacy. Instead of aiming for clarity or action, DataCrumb made space for noticing and questioning. This approach opens possibilities for designing interventions that work with, rather than resolve, the complexity of invisible digital systems. Future work may extend this by exploring long-term use, shared reflection, or hybrid forms that combine provocation with optional explanation.

\section*{Reproduction Note} \label{RP}
The interview questions, transcripts, design files, and source code are available publicly at \url{https://osf.io/hu98v/overview}.

%\begin{acks}
%\end{acks}

\bibliographystyle{ACM-Reference-Format}
\bibliography{references}
\appendix

\end{document}